\documentclass[aps,twocolumn,superscriptaddress,letterpaper]{revtex4}
\usepackage{graphicx,amssymb,bm,natbib,amsfonts,amsmath}
\usepackage{hyperref}

\newcommand {\qN}{{\bf q}$_N$}
\newcommand {\qone}{{\bf q}$_1$}
\newcommand {\qtwo}{{\bf q}$_2$}

\begin{document}

\title{Revealing Charge Density Wave Formation in the LaTe$_{2}$ System by Angle Resolved Photoemission Spectroscopy}

\author{D.R. Garcia}
\affiliation{Department of Physics, University of California,
Berkeley, CA 94720, USA}
\affiliation{Materials Sciences Division,
Lawrence Berkeley National Laboratory, Berkeley, CA 94720, USA}
\author{G.-H. Gweon$^*$}
\affiliation{Department of Physics, University of California,
Berkeley, CA 94720, USA}
\author{S.Y. Zhou}
\affiliation{Department of Physics, University of California,
Berkeley, CA 94720, USA}
\affiliation{Materials Sciences Division,
Lawrence Berkeley National Laboratory, Berkeley, CA 94720, USA}
\author{J. Graf}
\affiliation{Materials Sciences Division, Lawrence Berkeley National
Laboratory, Berkeley, CA 94720, USA}
\author{C.M. Jozwiak}
\affiliation{Department of Physics, University of California,
Berkeley, CA 94720, USA}
\author{M.H. Jung}
\affiliation{National Research Laboratory for Material Science, KBSI, Daejeon 305-333, South Korea}
\author{Y.S. Kwon}
\affiliation{Department of Physics, Sung Kyun Kwan University, Suwon 440-746, South Korea}
\author{A. Lanzara $^\dagger$}
\affiliation{Department of Physics, University of California,
Berkeley, CA 94720, USA}
\affiliation{Materials Sciences Division, Lawrence Berkeley National Laboratory, Berkeley, CA 94720, USA}

\date{\today}

\begin{abstract}
We present the first direct study of charge density wave (CDW) formation in quasi-2D single layer LaTe$_2$ using high-resolution angle resolved photoemission spectroscopy (ARPES) and low energy electron diffraction (LEED).  CDW formation is driven by Fermi surface (FS) nesting, however characterized by a surprisingly smaller gap ($\approx 50$meV) than seen in the double layer RTe$_3$ compounds, extending over the entire FS.  This establishes LaTe$_2$ as the first reported semiconducting 2D CDW system where the CDW phase is FS nesting driven.  In addition, the layer dependence of this phase in the tellurides and the possible transition from a stripe to a checkerboard phase is discussed.
\end{abstract}

\maketitle
The physics of the charge density wave (CDW) state remains among the most actively studied phenomena in solid-state physics due to its competition and even coexistence with superconductivity \cite{Nunezregueiro}-\cite{Jung1}, its potential role in the superconducting cuprate phase diagram \cite{Dung-Hai, Kyle}, and its important insights into electron-phonon physics.  The origin of a CDW is most commonly traced to a Fermi surface (FS) nesting, i.e. the matching of sections of FS to others by a single wave vector, {\bf q}$_N$.  For higher dimensional FSs the CDW phase tends to remain metallic, either due to imperfect nesting which leaves regions of the FS ungapped \cite{Gweon, Gweon2}, or to residual electron pockets formed by the CDW formation \cite{Brouet, Yokoya}.  This is in contrast to quasi-1D CDW systems, where a perfect nesting can be realized and the FS is fully gapped, explaining why all known quasi-1D CDW materials are semiconductors in the CDW phase \cite{Gruner, Gweon-JPCM, Nunezregueiro}. While we can find a few examples of non-metallic 2D CDW systems, the origin of the CDW phase is not due to true FS nesting but Mott physics \cite{Colonna, Kim}) or other non-nesting phenomena \cite{T.E.Kidd}.
Therefore, a natural question is whether there exists any proven instance of a 2D CDW system where the CDW phase is driven by FS nesting, yet non-metallic.  

The 2D CDW rare earth ditellluride system LaTe$_2$ \cite{DiMasi}-\cite{Shin}, formed by square tiled Te layers separated by RTe slabs (R = rare earth), is ideal to address this question, having been previously shown to be non-metallic \cite{Shin, Kwon}.  The CDW phase is well established, first by transmission electron microscopy (TEM) measurements which identified a modulation wave vector $0.5 {\bf a}^*$ \cite{DiMasi}, which we will refer to as \qone, and later by single crystal X-ray diffraction \cite{Stowe}, which proposed a $2\times2\times1$ superstructure.  Recently TEM measurements have reported another CDW vector with $\bf{q}$=$.6\bf{a}$$^*$+.2$\bf{b}$$^*$ \cite{Shin} which we will refer to as $\bf{q}_2$.  While ARPES has successfully studied other tellurides such as RTe$_3$, showing its CDW phase to be FS driven, only recently has it been used for the LaTe$_{2}$ system \cite{Shin}, although a complete study of the CDW phase is missing. 

	In this Letter, we present the first detailed ARPES study of the band structure and charge density wave formation in LaTe$_2$.  Like other tellurides, the CDW phase is driven by FS nesting, with $\bf{q}$$_{1}$=.53$\bf{a}$$^*$.  However it is characterized by a surprisingly small gap ($\approx 50$ meV measured from the leading edge) compared to RTe$_3$ compounds \cite {Gweon, Brouet, Ru-chemical-pressure} which extends over the {\em entire} FS.  No evidence has been observed for a corresponding FS nesting driven by the $\bf{q}$$_{2}$ vector, suggesting its origin might be imperfect stoichiometry in the crystal's chalcogenide planes as seen in the iso-structural selenides \cite {Grupe, Lee}.  These results suggest that LaTe$_2$ is the first proven example of a quasi-2D system with a CDW phase both FS driven and semiconducting as seen in 1D CDW materials.  

    ARPES and LEED data were taken on LaTe$_{2}$ single crystals at BL7.0.1 and BL10.0.1 of the Advanced Light Source using Scienta SES100 and R4000 analyzers.  A total energy resolution of $<$40 meV was used, with an angular resolution set to 0.35 degrees.  Samples were cleaved {\em in situ} with a base pressure better than $7\times 10^{-11}$ torr at low temperatures.  All APRES data was measured using 110eV photons, and the chemical potential, $\mu$, determined from gold foil with an uncertainty of $\pm$0.5meV.  

\begin{figure}
\includegraphics[width=7.5 cm]{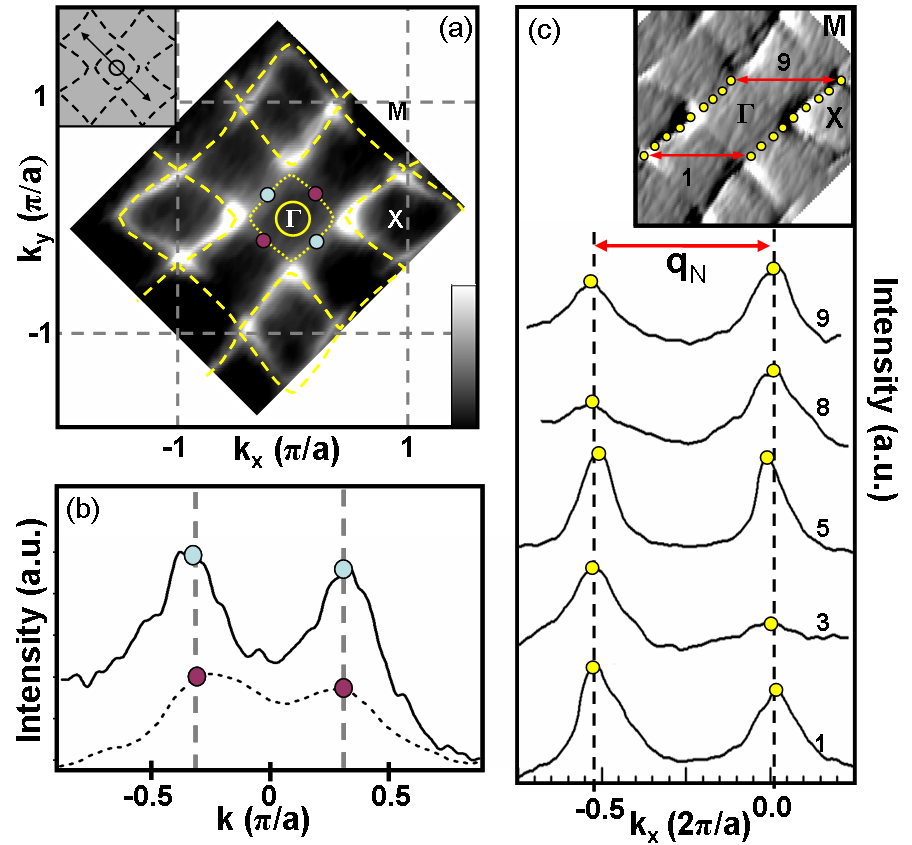} 
\caption{(color online)(a) Unsymmetrized FS map (integrated between +50 meV above to -100meV below the chemical potential, $\mu$) at T=180K.  White represents maximum intensity and black zero intensity with beam polarization displayed in the inset.  The non-CDW LDA band structure at E$_F$ \cite{Shim} is shown as yellow lines.  (b) MDCs through the inner diamond, along M-$\Gamma$-M directions.  (c) Raw MDCs spectra near $\mu$ for cuts parallel to the $\Gamma$-X direction (1 - 9) as shown in the inset of the same figure, where the first derivative of the FS is shown.  Each curve is shifted in k-space by a constant.} 
\end{figure}

	  Fig.~1a shows the constant binding energy band structure of LaTe$_2$ near $\mu$.  The experimental FS fits remarkably well with independent LDA calculation \cite{Shim} (yellow lines).  Along with the outer FS (dashed lines, ``outer contour''), of particular interest is a small square contour centered around the $\Gamma$ point (dotted lines, `inner diamond'), predicted by LDA but never resolved within this energy window \cite{Shin}.  Using photon energies closer to the Te 4p orbital binding energy ($\approx$103eV) appears to be crucial to resolving this feature.  The location of the inner diamond is best quantified from momentum distribution curves (MDCs; intensity vs.~momentum at constant energy) shown in panel b.  The importance of both the inner diamond and the stronger outer contour lies in the essential role they play in determining CDW formation as discussed later.  While theory and experiment agree well for these features, the small $\Gamma$ point electron pocket (yellow solid line) predicted by LDA is not resolved even over a range of 80-200eV in photon energy.

\begin{figure}
\includegraphics[width=9.0 cm]{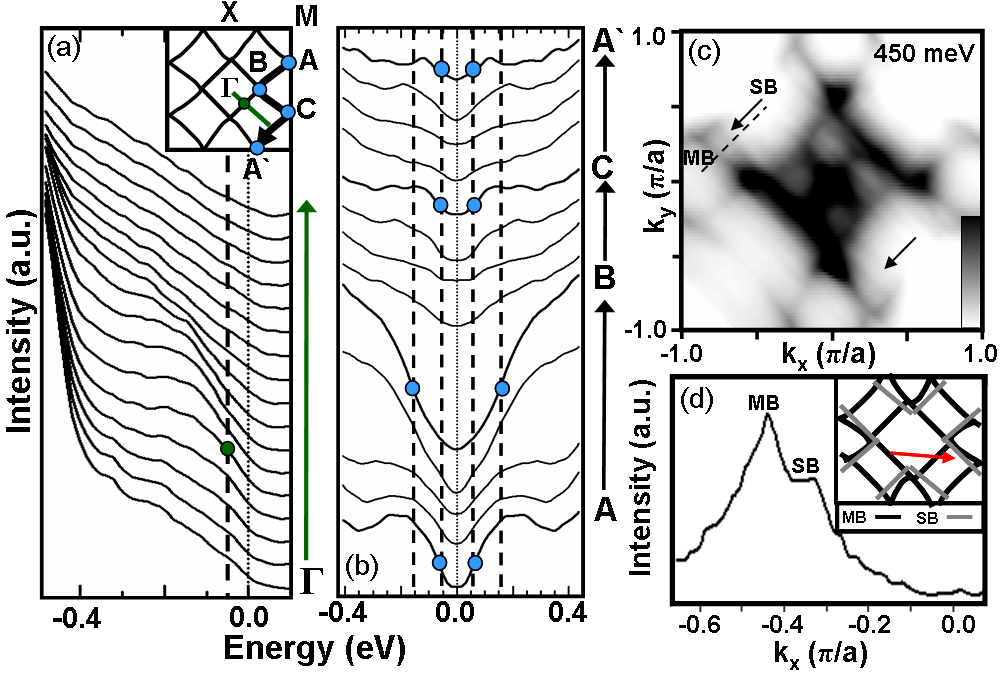}
\caption{(color online) (a) EDCs stack taken along the green line ($\Gamma$-M) in the inset.  (b) Symmetrized EDC stack at $k_F$ along one quarter of the outer band contour as indicated in the inset of 2a, with T=50K. (c) Unsymmetrized constant energy cut at 450meV showing evidence of shadow bands indicated by black arrows, where black represents maximum intensity and white zero intensity. (d) MDC cut taken along the dashed line in panel c showing two peaks associated with the main band (MB) and the shadow band (SB).  The inset illustrates how the shadow bands (gray line) arise by shifting the inner diamond by the nesting vector $\bf{q}$$_{N}$ (red arrow).}  
\end{figure}

This FS can be approximately represented by two perpendicular pairs of nearly 1D bands parallel to the $\Gamma$-M direction (see inset of Fig.~1c).  This gives rise to an almost perfectly nested FS, favoring a CDW with nesting vector parallel to the $\Gamma$-X direction (red arrows in Fig.~1c inset).  The nesting vector $\bf{q}$$_{N}$=.53$\bf{a}$$^*$ is obtained from the separation between the MDCs peaks shown in Fig.~1c (yellow circles), taken for different cuts (1 to 9) along the $\Gamma$-X direction and is consistent with prior scattering work.  This nesting is nearly perfect, with a variation in the peak to peak distance of $\approx$2\% as compared to $\approx$20\% in tritelluride \cite{Gweon}. Finally, since the FS pattern is four-fold symmetric, the MDCs shown suggest that $\bf{q}$$_{N}$ also nests the entire FS, explaining the \qone~observed by TEM\@.

\begin{figure*}
\includegraphics[width=18.0 cm]{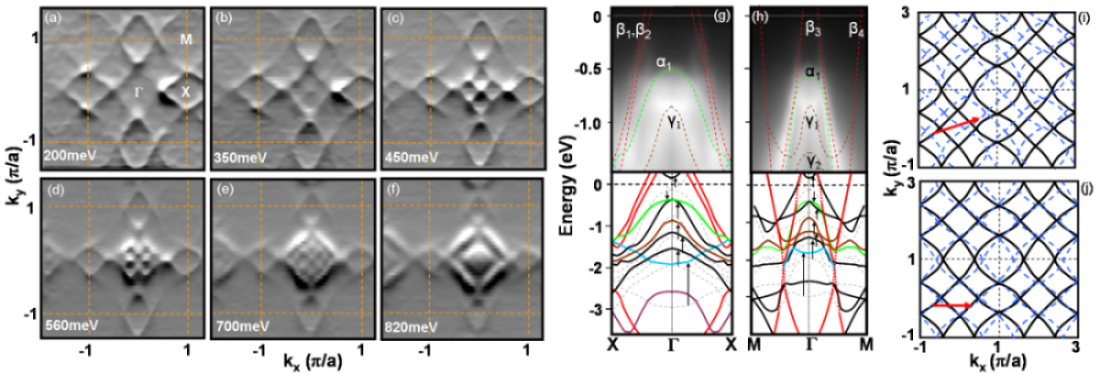}
\caption{(color online) (a-f) First derivative with respect to k$_y$ of the unsymmetrized constant energy maps over energies 200 to 820meV at T=180K.  (g),(h) Image plots taken along the high symmetry directions: X-$\Gamma$-X and M-$\Gamma$-M, respectively.  The lower panels show the LDA band structure.  The suggested renormalization of the LaTe block bands with respect to their original energies (gray dashed lines) is shown by black arrows.  (i) Cartoon illustrating the experimental FS structure of Fig. 1a (black curves) shifted by \qtwo~(red arrow and blue dashed lines) indicating a lack of any obvious nesting by this wavevector for the data. This is compared to (j) where the band structure is shifted by \qone~and shows the nestable regions examined in Fig.~1c}
\end{figure*}

The momentum dependent energy gap can be measured from the leading edge position of the energy distribution curves (EDCs; intensity vs.~energy at constant momentum) at k$_F$, a standard ARPES procedure.  Fig.~2a shows a series of EDCs along the $\Gamma$-M direction (green line in the inset) over the inner diamond contour.  We resolved a small peak in the EDC, which disperses toward $\mu$, yet never crosses it, and eventually recedes to higher binding energy.  This allows us to estimate a leading edge midpoint gap of 50 $\pm$ 10meV (green circle), suggesting $\Delta \approx 100$ meV\@.  This is the first characterization of the inner diamond gap as prior studies had not resolved this diamond and assigned $\Delta \approx 600$ meV \cite {Shin}.

For the outer FS, Fig.~2b shows a stack of EDC spectra at k$_F$ along the outer contour, from A to A' (inset of Fig.~2a).  To aid in identifying FS gapping, these EDCs have been symmetrized to remove contributions by the Fermi function, a well known technique detailed elsewhere \cite{Kanigel}.  The midpoints of the leading edge show that the majority of the spectra are also gapped by $\approx 50$ $\pm$ 14meV, similar to the inner diamond (blue circles at high symmetry points), although deviations from this are observed.  This gap behavior is four-fold symmetric throughout the entire BZ.  Surprisingly, the gap seems to anomalously increase near the B points, closest to $\Gamma$ point, along the contour.  Although a more complicated CDW origin for this anomalous gapping is possible, we propose a simpler explanation.  Two bands are predicted to exist near the B point, one due to the inner diamond and other from the outer FS.  The anomalous increase of the gap can then be explained as a shift of the relative spectral weight between these two bands.  

Finally, the observation of shadow bands corresponding to our $\bf{q}$$_{N}$ further supports it as the CDW vector of the system.  In the constant energy map shown in Fig.~2c (black arrows), we find two peaks associated with the main and shadow bands in the MDC spectra (Fig.~2d) for a cut along the dashed line in Fig.~2c.  The inset illustrates how these twin peaks are produced by shifting the inner diamond main band by \qN~of Fig.~1c.  In summary, CDW formation with $\bf{q}$$_{1}$ is induced by a FS nesting instability, which leads to the opening of a CDW gap along the FS where the majority of the contour is gapped on the order of 50meV measured from the leading edge.  The fact that this gap persists over the {\em entire} FS is consistent with semiconducting properties \cite{Kwon,Shin}.  

In contrast, it is hard to find any strong ARPES signature of the second CDW order $\bf{q}_2$\cite{Shin}.  Specifically, we could not identify any FS band structure which could be nested by this wavevector (illustrated by Fig.~3i). Fig.~3 shows constant energy maps at higher binding energy (in the range between 200 and 820meV).  By increasing the binding energy, we can easily follow the expected evolution of the outer Fermi surface contour (Figs.~3a,b).  At $\approx 400$meV, we observe the onset of a far more complicated structure inside the inner diamond, not readily explained by LDA (panels c-f).  One possibility for these structures is that they reflect an ordered state, such as a CDW phase, related to the $\bf{q}_2$ vector.  Unfortunately, this did not pass a careful examination, since no such nesting wavevector could be identified.

    A simpler explanation is that these patterns result from the crossing of different bands within this energy window.  Having done a detailed photon energy study, we observe that these bands quickly disappear as we tune the energy away from the La 4d$_{3/2}$ adsorption edge ($\approx$105eV), suggesting their La character.  We then compared the LDA band structure with our experimental dispersions along the two high symmetry directions, M-$\Gamma$-M and X-$\Gamma$-X (Figs.~3g,h).  Experiment shows a large number of bands at higher binding energy, between 0.4 to 0.8eV.  Addressed in more detail elsewhere \cite{ICESS}, this points to a renormalization toward lower binding energy for the LaTe block bands (e.g. $\gamma$$_1$ and $\gamma$$_2$), not uncommon for LDA, as indicated by the arrows in the LDA model.  In addition, the $\Gamma$ point electron pocket from Fig.~1a and the $\alpha$$_{1}$ band, which both arise from the LaTe block, are pushed apart by a greater energy difference than LDA, indicating a different hybridization strength.  This explains the absence of the electron pocket in Fig.~1a, as being pushed above E$_F$.  These two renormalizations then shift the LaTe block bands into the energy region of Figs.~3c-f leading to these remarkable k-space patterns.  Nevertheless, a better theoretical understanding of our data is still needed.  Additionally, the strong increase in band intensity near the $\Gamma$ point (Fig.~3) overpowers the weaker low energy bands which the CDW actually gaps, and might be responsible for the earlier tunneling report of a much larger CDW gap, $\Delta = 0.45$eV \cite{M.H.Jung}.
    
	To better understand the \qtwo~vector reported by TEM, Fig.~4a shows LEED taken on the sample surface.  As with TEM, the main Bragg spots obey an h + k = even condition \cite{DiMasi,Shin}.  In addition to these main spots, satellite peaks appear at wave vectors related to approximately $\bf{q}$=.6$\bf{a}$$^*$+.2$\bf{b}$$^*$ from the main Bragg spots.  Also, unlike the TEM work, the LEED pattern breaks mirror symmetry, revealing only half the satellite peaks. This unique symmetry breaking associated with \qtwo~also appears in our ARPES data. Fig.~4b shows a constant energy map centered around 130meV in binding energy.  We identify features near the X points (red arrows in panel b) which are dispersive and gradually disappear with time, thus suggesting surface states.  Like the LEED pattern, these features are four-fold symmetric but break mirror symmetry.  We also note that no evidence of a surface 2$\times$2$\times$1 superstructure was observed in our LEED studies.  As mentioned, the comparative weakness of the shadow bands suggests that the modulation in electron density is very small and explains its absence in LEED which only discerns total electron density. From these results, we propose that the \qtwo~modulation seen in LEED and its mirror symmetric pair both exist as superstructure domains in the bulk and are caused by an ordered-defect in the crystal's square chalcogenide planes from imperfect stoichiometry.  Such defect ordering is expected and observed in the isostructural LaSe$_{1.9}$ \cite{Grupe,Lee}.  The superposition of mirror symmetric superstructure domains explains the symmetry of TEM results.  Yet at the surface, only the superstructure of a single Te plane is observed which breaks mirror symmetry, dominates the LEED pattern, and causes surface states reflecting this broken symmetry seen in ARPES data.  We also speculate that defects mainly affect high energy states by trapping charges, not the states near $\mu$.

\begin{figure}
\includegraphics[width=8.0 cm]{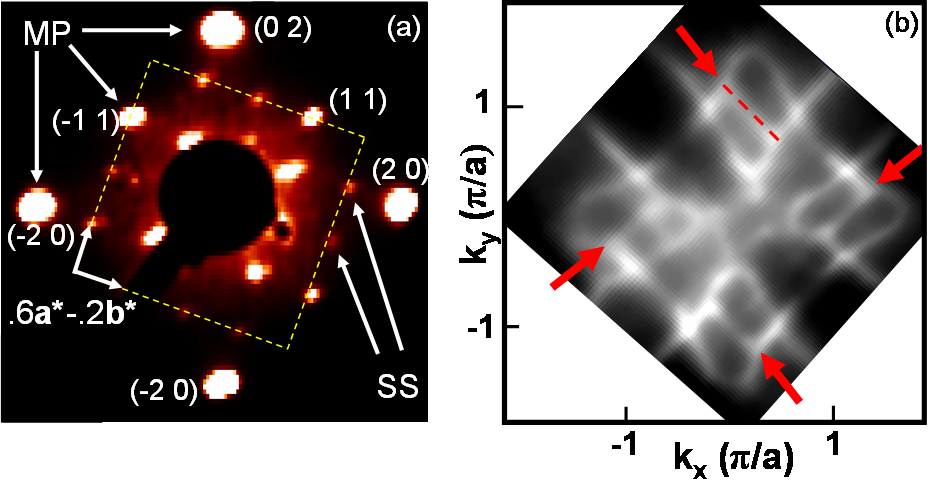}
\caption{(color online) (a) LEED measurements of sample surface using 95eV electrons at 160K. The $\bf{q}$=$.6\bf{a}$$^*$-.2$\bf{b}$$^*$ superstructure is indicated by white arrows. (b) Unsymmetrized constant energy map integrated over 100-160meV at 50K.  The anomalous bands are indicated by red arrows.}
\end{figure}

Comparing the single layer LaTe$_2$ and double layer RTe$_3$, this study suggests that the FS driven nature of the $\bf{q}$$_{CDW}$ does not change between the two compounds, although the FS nesting in LaTe$_2$ appears more perfect than the tritellurides \cite{Gweon,Brouet}.  Secondly, the CDW gap is several times smaller that of tritellurides \cite {Gweon,Brouet}. Finally, recent theory on the rare earth tellurides, RTe$_n$, suggests two possible ordered phases: a stripe phase and a checkerboard phase \cite{Yao}.  From our results, the single layered compounds might fall closer to the CDW checkerboard phase than RTe$_3$ systems. It also seems possible that if the CDW interaction is reduced, e.g.~by applying pressure \cite {Nunezregueiro}, the checkerboard pattern may emerge. 
To conclude, we have presented the first direct study of CDW formation in single layer LaTe$_{2}$.  CDW formation in this material is a FS driven phenomena characterized by a CDW gap which opens over the entire FS despite its 2D nature.  These results establish LaTe$_2$ as the first proven instance of a quasi-2D CDW material whose CDW phase is both driven by FS nesting and semiconducting.  The large changes as the number of Te layers reduces from two (in RTe$_3$) to one (in RTe$_2$), such as a large decrease of CDW gap size, a large increase of gap isotropy, and the emergence of superconductivity (CeTe$_{1.82}$ \cite{Jung1}), are all intriguing properties.  Understanding these changes may shed light on other correlated electron problems such as high temperature superconductivity, where the number of layers is already known to play a crucial role in determining the pseudogap formation temperature.

We thank A. Castro-Neto, K. McElroy, and A. Bill for useful discussions. This work was supported by the Director, Office of Science, Office of Basic Energy Sciences, Division of Materials Sciences and Engineering of the U.S DOE under Contract No.~DEAC03-76SF00098 and by the NSF through Grant No. DMR03-49361.  The ALS is supported by the Director, Office of Science, Office of Basic Energy Sciences of the U.S. DOE under Contract No.\ DE-AC02-05CH11231. 

$^\dagger$ Electronic address: alanzara@lbl.gov

$^*$ Current address: Department of Physics, University of California, Santa Cruz, CA, 95060


\begin {thebibliography} {99}

\bibitem{Nunezregueiro} M. Nunez-Regueiro et al., {\em Synth. Met.} {\bf56}, 2653 (1993).
\bibitem{Morris} R.C. Morris, Phys.\ Rev.\ Letters {\bf34}, 1164 (1975).
\bibitem{Singh} Y. Singh et al., Phys.\ Rev.\ B {\bf72}, 45106 (2005).
\bibitem{Fang} L. Fang et al., Phys.\ Rev.\ B {\bf72}, 14534 (2005).
\bibitem{Morosan} E. Morosan et al., {\em Nature Physics} {\bf2}, 544 (2006).
\bibitem{Jung1} M. H. Jung et al., Phys.\ Rev.\ B {\bf67}, 212504 (2003).
\bibitem{Dung-Hai} Jian-Xin Li et al., Phys.\ Rev.\ B {\bf74}, 184515 (2006).
\bibitem{Kyle} K. McElroy et al., Phys.\ Rev.\ Lett. {\bf94}, 197005 (2005).
\bibitem{Gweon} G.-H. Gweon et al., Phys.\ Rev.\ Lett. {\bf81}, 886 (1998).
\bibitem{Gweon2} G.-H. Gweon et al., Phys.\ Rev.\ B {\bf55}, R13353 (1997).
\bibitem{Brouet} V. Brouet et al., Phys.\ Rev.\ Lett. {\bf93}, 126405 (2004).
\bibitem{Yokoya} T. Yokoya et al., Science {\bf294}, 2518-2520 (2001).
\bibitem{Gweon-JPCM} G.-H. Gweon et al., J.~Phys.~Cond.~Matt.~{\bf 8}, 9923 (1996).
\bibitem{Gruner} G. Gr\"uner, {\em Density Waves in Solids} Addison-Wesley, Reading, Massachusetts, (1994).
\bibitem{Colonna} S. Colonna et al., Phys.\ Rev.\ Letters {\bf94}, 36405 (2005).
\bibitem{Kim} J.J. Kim et al., Phys.\ Rev.\ Letters {\bf73}, 2103 (1994).
\bibitem{T.E.Kidd} T.E. Kidd et al., Phys.\ Rev.\ Letters {\bf88}, 226402 (2002).
\bibitem{DiMasi} E. DiMasi et al., Phys.\ Rev.\ B {\bf54}, 13587 (1996).
\bibitem{Stowe} K. St\"owe, J. Solid State Chem. {\bf149}, 155 (2000).
\bibitem{Shin} K.Y. Shin et al., Phys.\ Rev.\ B {\bf72}, 85132 (2005).
\bibitem{Kwon} Y.S. Kwon and B.H. Min, Physica B {\bf281-282}, 120-121 (2000).
\bibitem {Ru-chemical-pressure} N. Ru et al., http://arXiv.org/cond-mat/0610319.
\bibitem{Grupe} M. Grupe and W. Urland, J. Less Com. Met. {\bf170}, 271 (1991).
\bibitem{Lee} S. Lee and B. Foran, J.~Am.~Chem.~Soc. {\bf116}, 154 (1994).
\bibitem{Shim} J.H. Shim, et al., Phys.\ Rev.\ Letters {\bf93}, 156406 (2004).
\bibitem{Kanigel} A. Kanigel et al., {\em Nature Physics} {\bf2}, 447-451 (2006).
\bibitem{ICESS} D.R. Garcia et al., J.~Elec.~Spec.~Rel.~Phen. {\bf156-158}, 58-63 (2007).
\bibitem{M.H.Jung} M.-H. Jung et al., Phys.\ Rev.\ B {\bf63}, 035101 (2000).
\bibitem{Yao} H. Yao et al., Phys.\ Rev.\ B {\bf74}, 245126 (2006).

\end {thebibliography}
\end{document}